# High-mobility two-dimensional carriers from surface Fermi arcs in magnetic Weyl semimetal films

Shingo Kaneta-Takada,[1,2,*] Yuki K. Wakabayashi,[1,*,†] Yoshiharu Krockenberger,[1] Toshihiro Nomura,[3] Yoshimitsu Kohama,[3] Sergey A. Nikolaev,[4,5] Hena Das,[4,5] Hiroshi Irie,[1] Kosuke Takiguchi,[1,2] Shinobu Ohya,[2,6] Masaaki Tanaka,[2,7] Yoshitaka Taniyasu,[1] and Hideki Yamamoto[1]

[1]*NTT Basic Research Laboratories, NTT Corporation, Atsugi, Kanagawa 243-0198, Japan*
[2]*Department of Electrical Engineering and Information Systems, The University of Tokyo, Bunkyo, Tokyo 113-8656, Japan*
[3]*Institute for Solid State Physics, The University of Tokyo, Kashiwa, Chiba 277-8581, Japan*
[4]*Laboratory for Materials and Structures, Tokyo Institute of Technology, 4259 Nagatsuta, Midori-ku, Yokohama, Kanagawa 226-8503, Japan*
[5]*Tokyo Tech World Research Hub Initiative (WRHI), Institute of Innovative Research, Tokyo Institute of Technology, 4259 Nagatsuta, Midori-ku, Yokohama, Kanagawa 226-8503, Japan*
[6]*Institute of Engineering Innovation, The University of Tokyo, Bunkyo, Tokyo 113-8656, Japan*
[7]*Center for Spintronics Research Network (CSRN), The University of Tokyo, Bunkyo, Tokyo 113-8656, Japan*

[*]These authors contributed equally to this work.
[†]Corresponding author: yuuki.wakabayashi.we@hco.ntt.co.jp

## Abstract

High-mobility two-dimensional carriers originating from surface Fermi arcs in magnetic Weyl semimetals are highly desired for accessing exotic quantum transport phenomena and for topological electronics applications. Here, we demonstrate high-mobility two-dimensional carriers that show quantum oscillations in magnetic Weyl semimetal $SrRuO_3$ epitaxial films by systematic angle-dependent, high-magnetic field magnetotransport experiments. The exceptionally high-quality $SrRuO_3$ films were grown by state-of-the-art oxide thin film growth technologies driven by machine learning algorithm. The quantum oscillations for the 10-nm $SrRuO_3$ film show a high quantum mobility of $3.5 \times 10^3$ cm$^2$/Vs, a light cyclotron mass, and two-dimensional angular dependence, which can be attributed to the surface Fermi arcs. The linear thickness dependence of the phase shift of the quantum oscillations provides evidence for the non-trivial nature of the quantum oscillations mediated by the surface Fermi arcs. In addition, at low temperatures and under magnetic fields of up to 52 T, the quantum limit of $SrRuO_3$ manifests the chiral anomaly of the Weyl nodes. Emergence of the hitherto hidden two-dimensional Weyl states in a ferromagnetic oxide pave the way to explore novel quantum transport phenomena for topological oxide electronics.

# I. INTRODUCTION

High-mobility two-dimensional carriers were first realized at semiconductor surfaces and interfaces, where they form the basis of the integer and fractional quantum Hall effects [1–4], two-dimensional superconductivity [5], and practical applications such as high-electron-mobility transistors [6,7]. The two-dimensional carriers originating from the surface Fermi arcs or the quantum confinement of bulk three-dimensional Dirac fermions are reported for the Dirac semimetal $Cd_3As_2$ [8–15]. The high-mobility two-dimensional carriers are also pursued in Weyl semimetals and magnetic Weyl semimetals. Examples of Weyl semimetals and magnetic Weyl semimetals are TaAs [16–19], $Na_3Bi$ [20,21], NbAs [22,23], $Co_3Sn_2S_2$ [24,25], $Co_2MnGa$ [26], and $SrRuO_3$ [27]. Since these states in topological semimetals are predicted to be robust against perturbations [10,14,28,29], they are of both scientific and technological interests, with potential for use in high-performance electronics, spintronics, and quantum computing. In particular, Weyl fermions in magnetic Weyl semimetals are thought to be more suitable for spintronic applications [25,30,31]. Since the distribution of Weyl nodes in magnetic Weyl semimetals is determined by the spin texture [27,32], the high-mobility two-dimensional carriers originating from the Weyl nodes in magnetic Weyl semimetals are expected to be controllable by magnetization switching [33] in addition to the electric field [6,13]. To detect these high-mobility two-dimensional carriers in transport measurements, high-quality epitaxial Weyl semimetal thin films, in which the contributions of surface states and/or quantum-confined bulk states are prominent, are needed. Conversely, the experimental detection of these Weyl states in transport have remained elusive due to the lack of high-quality epitaxial Weyl semimetal films. Revealing the quantum transport properties of high-mobility two-dimensional carriers in magnetic Weyl semimetals is urgently required for a deep understanding of magnetic Weyl semimetals and demonstrating the relevance of Weyl fermions to spintronic and electronic applications.

The recent observation of the Weyl fermions in the itinerant $4d$ ferromagnetic perovskite $SrRuO_3$ [27,32,34], which is widely used as an epitaxial conducting layer in oxide heterostructures [35–44], points to this material's being an appropriate platform to integrate two emerging fields: topology in condensed matter and oxide electronics. Since the quantum transport of Weyl fermions in $SrRuO_3$ has been reported in high-quality epitaxial films [27], $SrRuO_3$ provides a promising opportunity to realize high-mobility two-dimensional carriers in magnetic Weyl semimetals. However, their dimensional character remains unsolved due to the lack of angle-dependent quantum oscillations. Since the amplitude of the quantum oscillation becomes small for angles close to the in-plane direction of the film, high magnetic field transport measurements are indispensable to scrutinize the angular dependence of the quantum oscillations. Furthermore, high magnetic field transport measurements will simultaneously provide information regarding the chiral anomaly in the quantum limit of Weyl fermions in $SrRuO_3$.

In this article, we present a systematic study of the angle-dependent magnetotransport, including quantum oscillations, up to 52 T in ultrahigh-quality $SrRuO_3$ thin films with thicknesses of 10, 20, 40, 60, and 63 nm grown by machine-learning-assisted molecular beam epitaxy (MBE) [45]. The quantum oscillations in resistivity [i.e., Shubnikov-de Haas (SdH) oscillations] for the 10-nm film show the light cyclotron mass of $0.25m_0$ ($m_0$: free electron mass in a vacuum), high quantum mobility of $3.5 \times 10^3$ cm$^2$/Vs, and two-dimensional angular dependence, confirming the existence of high-mobility two-dimensional carriers in SrRuO$_3$. The quantum oscillations for the 63-nm film, whose thickness is larger than the Fermi wavelength of the high-mobility carriers (~22.4 nm), also show the two-dimensional angular dependence, suggesting that the high-mobility two-dimensional carriers come from the surface Fermi arcs, not the quantum confinement effect. The thickness-dependent phase shift establishes the non-trivial character of the quantum oscillations mediated from the surface Fermi arcs: The phase shift of the quantum oscillations shows a linear thickness dependence. We also observed the saturation of the chiral-anomaly-induced negative MR in the quantum limit, which had never been observed in magnetic Weyl semimetals. These findings of the high-mobility two-dimensional carriers and chiral anomaly in the quantum limit in an epitaxial ferromagnetic oxide provide an intriguing platform for topological oxide electronics and will stimulate further investigation of exotic quantum transport phenomena in epitaxial heterostructures.

## II. EXPERIMENTS
### A. Sample preparation

We grew high-quality epitaxial SrRuO$_3$ films with thicknesses $t$ = 10, 20, 40, 60, and 63 nm on (001) SrTiO$_3$ substrates [Fig. 1(a)] in a custom-designed MBE setup equipped with multiple e-beam evaporators for Sr and Ru [27,45]. The SrRuO$_3$ films with $t$ = 10-40 nm and 63 nm were prepared under the same growth conditions as those in Ref. [27] and Ref. [47], respectively. The growth parameters, such as the growth temperature, supplied Ru/Sr flux ratio, and oxidation strength, were optimized by Bayesian optimization, a machine learning technique for parameter optimization [45,48,49], with which we achieved a high residual resistivity ratio (RRR) of 19.2 and 81.0 with $t$ = 10 and 63 nm, respectively. We precisely controlled the elemental fluxes, even for elements with high melting points, e.g., Ru (2334°C), by monitoring the flux rates with an electron-impact-emission-spectroscopy sensor, which was fed back to the power supplies for the e-beam evaporators. The stoichiometric Sr/Ru ratio of a SrRuO$_3$ film was confirmed using energy dispersive x-ray spectroscopy [50]. The growth rate of 1.05 Å/s was deduced from the thickness calibration of a thick (63 nm) SrRuO$_3$ film using STEM. Epitaxial growth of the high-quality single-crystalline SrRuO$_3$ film with an abrupt substrate/film interface is seen in EELS-STEM images [Fig. 1(b)]. In Refs. [27] and [47], the high crystalline quality and large coherent volume of the epitaxial single-crystalline films were also confirmed by comprehensive crystallographic analyses using reflection high-energy electron diffraction and $\theta$–$2\theta$ x-ray diffraction. Further information about the MBE setup, preparation of the substrates, and sample properties are described elsewhere [27,50–53].

### B. Magnetotransport measurements

We fabricated 200 × 350 μm$^2$ Hall bar structures [Fig. 1(c)] by photolithography and Ar ion milling. We deposited Ag electrodes on a SrRuO$_3$ surface before making the Hall bar structure. Magnetotransport up to 14 T was measured in the PPMS sample chamber equipped with a rotating sample stage. High-field magnetotransport measurements up to 52 T and 40 T were performed using a non-destructive mid-pulse magnet with a pulse duration of 36 ms and long-pulse magnet with a pulse duration of 1.2 s, respectively, at the International MegaGauss Science Laboratory at the Institute for Solid State Physics, the University of Tokyo [54].

### C. First-principles calculations

Electronic structure calculations were performed within density functional theory by using generalized gradient approximation [55] for the exchange correlation functional in the projector-augmented plane wave formalism [56] as implemented in the Vienna Ab-initio Simulation Package [57]. The energy cutoff was set to 500 eV, the Brillouin zone was sampled by an 8×8×6 Monkhorst-Pack mesh [58], and the convergence criterion for electronic density was put to 10$^{-8}$ eV. Experimental crystal structure of orthorhombic SrRuO$_3$ was adopted for all calculations (the *Pbnm* space group, a = 5.567 Å, b = 5.5304 Å, c = 7.8446 Å [39]). The calculations were performed for the ferromagnetic state with spin-orbit coupling. The effect of electronic correlations in the Ru 4*d* shell was considered by using the rotationally invariant GGA + *U* scheme [59] with *U* = 2.6 eV and *J* = 0.6 eV.

For numerical identification of the Weyl nodes, the electronic band structure was interpolated with the maximally localized Wannier functions by projecting the bands near the Fermi level onto the Ru *d* atomic orbitals, as implemented in the wannier90 package [60]. Band crossings in the reciprocal space were calculated by steepest-descent optimization of the gap function, $\Delta = (E_{n+1,\mathbf{k}} - E_{n,\mathbf{k}})^2$, on a uniform mesh for the Brillouin zone up to $25 \times 25 \times 25$ [61]. The two bands are considered degenerate when the gap is below the threshold of 10$^{-5}$ eV. The results of electronic structure calculations and the full set of the Weyl points in the vicinity of the Fermi level were reported in our previous study [27].

### D. Data pretreatment and Fourier transformation analysis for SdH oscillations

We subtracted the background of the conductivity $\sigma_{xx}$ from the raw data (see the Sec. I of the Supplemental Material [62]) using a polynomial function up to the eighth order and extracted the oscillation components. For the Fourier transformation, we interpolated the data in Figs. 5(a), 6(a), 7(a), 9, and S2(a) [62] to prepare equally spaced *x*-axis (1/*B*) points. Then, we multiplied the data with a Hanning window function to obtain the periodicity of the experimental data. Finally, we conducted a fast Fourier transform (FFT) on the data. The FFT frequencies *F* are determined from the peak positions of the FFT spectra obtained by the above procedure. The error bars of the FFT frequencies are defined as the full width at half maximum of the FFT peaks.

### III. RESULTS AND DISCUSSION
#### A. High-quality SrRuO$_3$ thin films and thickness-dependent magnetotransport

We performed magnetotransport measurements on the Hall bar devices of the SrRuO$_3$ films [Fig. 1(c)] with $t$ = 10 and 63 nm using a DynaCool physical property measurement system (PPMS). The temperature $T$ dependence of the longitudinal resistivity $\rho_{xx}$ of the SrRuO$_3$ films is shown in Fig. 2(a). The $\rho_{xx}$ of the films with $t$ = 10 and 63 nm decreased with decreasing temperature, indicating that these films are metallic over the whole temperature range. The $\rho_{xx}$ vs. $T$ curves show clear kinks [arrows in Fig. 2(a)], at which the ferromagnetic transition occurs, and spin-dependent scattering is suppressed [39]. The magnetization measurement for the SrRuO$_3$ film with $t$ = 63 nm at $T$ = 10 K shows a typical ferromagnetic hysteresis loop [left inset in Fig. 2(a)]. As shown in the right inset in Fig. 2(a), below 20 K, the SrRuO$_3$ films showed a $T^2$ scattering rate ($\rho \propto T^2$) that is expected for a Fermi liquid, in which electron-electron scattering dominates the transport and carriers are described as Landau quasiparticles [27,39,63,64]. Although the RRR decreases from 81.0 to 19.2 with decreasing $t$ from 63 to 10 nm, the RRR of the 10-nm film is high enough to observe the intrinsic quantum transport of Weyl fermions in SrRuO$_3$ [27,47]. The availability of such high-quality thin films allows to obtain new insights into the dimensionality of the Weyl fermions in SrRuO$_3$ *via* measurements of angle-dependent magnetotransport properties, as will be described in the Sec. III C. The larger residual resistivity $\rho_{Res}$ at 2 K for the SrRuO$_3$ film with $t$ = 10 nm suggests a disorder near the interface between SrRuO$_3$ and the SrTiO$_3$ substrate [47]. Figure 2(b) shows the MR (($\rho_{xx}(B)-\rho_{xx}(0\ T))/\rho_{xx}(0\ T)$) for the SrRuO$_3$ films with a magnetic field $B$ applied in the out-of-plane [001] direction of the SrTiO$_3$ substrate at 2 K. Importantly, the linear positive MR at 2 K showed no signature of saturation even up to 14 T, which is commonly seen in Weyl semimetals [9,19,65–68] and is thought to stem from the linear energy dispersion of Weyl nodes [69–71]. In the Fermi liquid temperature range ($T$ < 20 K), quantum lifetimes long enough to observe quantum oscillations were achieved, as evidenced by the observation of the SdH oscillations in both MR and $\rho_{xy}$ [Figs. 2(b) and 2(c)]. As will be described in the Sec. III C, the main component of the SdH oscillations for both films have the frequency $F \sim 25$ T, which was interpreted to be of the three-dimensional Weyl fermions with high mobility and light cyclotron mass in our previous study [27]. For clarifying the dimensionality of the SdH oscillations, we investigated the angular dependence of the magnetotransport properties in the SrRuO$_3$ films. The angle-dependent magnetotransport measurements also provide experimental evidence of the chiral anomaly that Weyl fermions in SrRuO$_3$ show.

#### B. Chiral anomaly in the quantum limit
We carried out the angle-dependent magnetotransport measurements for the SrRuO$_3$ film with $t = 10$ nm (Fig. 3). Fig. 3(a) shows the angular dependence of MR ($\rho_{xx}(B)-\rho_{xx}(0\ T))/\rho_{xx}(0\ T)$ for the SrRuO$_3$ film with $t = 10$ nm at 2 K measured by the PPMS (up to 14 T). Here, $B$ is rotated from the out-of-plane [001] direction of the SrTiO$_3$ substrate ($\theta = 90°$) to the in-plane direction parallel to the current ($\theta = 0°$). The rotation angle $\theta$ is defined in the inset of Fig. 3(a). As has been already described in the Sec. III A, the unsaturated linear positive MR is observed when $B$ is applied perpendicular to the current $I$ ($\theta = 90°$); note that quantum oscillations are superimposed onto the data. With decreasing $\theta$, the sign of the MR changes from positive to negative, and the negative MR ratio is enhanced. The negative MR at $\theta \leq 30°$ becomes linear above 5 T [Fig. 3(a)]. In ferromagnetic SrRuO$_3$, the time-reversal-symmetry breaking lifts the spin degeneracy and leads to linear crossing of non-degenerate bands at many $k$ points, resulting in a pair of Weyl nodes with opposite chiralities (L and R) [Fig. 3(c)] [27,32]. In the presence of a magnetic field, the Landau quantization of a pair of Weyl nodes with opposite chiralities occurs [Figs. 3(d) and 3(e)]. In Figs. 3(d) and 3(e), $B_{QL}$ represents the magnetic field at which the quantum limit is reached and all the Weyl fermions occupy the zeroth Landau levels. When $B < B_{QL}$ [Fig. 3(d)], nonorthogonal electric and magnetic fields ($\boldsymbol{E}\cdot\boldsymbol{B} \neq 0$) lead to the chiral charge transfer between the two Landau levels with opposite chiralities [72,73]. In the weak-$B$ limit ($B < B_{QL}$) [Fig. 3(d)], theoretical calculations based on the semiclassical Boltzmann kinetic equation predict that time-reversal-symmetry-breaking Weyl semimetals (magnetic Weyl semimetals) show a negative MR that is linear in the projection component of $B$ in the direction of $I$ [65,73,74], in contrast to the quadratic dependence expected for space-inversion-symmetry breaking Weyl semimetals [19,21,75]. Thus, the linear negative MR observed for $\theta \leq 30°$ and $5 \leq B \leq 14$ T when rotating $B$ from orthogonal to parallel to $I$ [Fig. 3(a)] is consistent with a chiral anomaly in magnetic Weyl semimetals [27].

To verify the negative MR stemming from the chiral anomaly, we carried out high-field angle-dependent magnetotransport measurements using the mid-pulse magnet up to 52 T. This is of importance because the negative MR at low field in Weyl semimetals can be also brought about by experimental artifacts and impurities [75–77]. Therefore, observing the behavior of the negative MR in the quantum limit (Fig. 3e), where the chiral-anomaly-induced negative MR is predicted to saturate [21,75,78], is necessary. Figure 3(b) shows the angular dependence of the high-field magnetotransport for $t = 10$ nm at 0.7 K with $B$ rotated from the out-of-plane [001] direction of the SrTiO$_3$ substrate to the in-plane direction parallel to the current. As in the case of the PPMS measurements up to 14 T [Fig. 3(a)], the sign of the MR at high fields changes from positive to negative with decreasing $\theta$ [Fig. 3(b)]. Importantly, the negative MR with $\theta = 5.3°$ saturates above 30 T [Fig. 3(b)], confirming that the negative MR originates from the chiral anomaly. In the quantum limit [Fig. 3(e)], the conductivity when $B$ is parallel to $I$ is expressed by

$$\sigma_{xx} = N_W \frac{e^2 v_F}{4\pi \hbar l_B^2} \tau_{inter}(B), \tag{1}$$

where $N_W$ is the number of Weyl points in the Brillouin zone, $v_F$ is the Fermi velocity,

$l_B = \sqrt{(h/eB)}$ is the magnetic length, and $\tau_{\text{inter}}(B)$ is the field-dependent internodal scattering time [72,78]. In the quantum limit of Weyl semimetals, the scattering rate $1/\tau_{\text{inter}}(B)$ is predicted to increase roughly in proportion to $B$ under the assumption of short-range impurity scattering, and the scattering factor cancels the factor of $l_B^2$ in eq. (1), resulting in the $B$-independent conductivity [75,78,79]. Indeed, the saturation of the chiral-anomaly-induced negative MR above the quantum limit ($B > B_{\text{QL}}$) has been observed in typical Weyl semimetals Na$_3$Bi [21] and TaAs [75] as a unique property of Weyl fermions, although it has not been observed in magnetic Weyl semimetals. Nonetheless, our observation of the saturation of the negative MR when $B$ is parallel to $I$ strongly indicates the existence of the chiral anomaly in the magnetic Weyl semimetal SrRuO$_3$.

The tilting of the Weyl nodes is an important feature that characterizes Weyl semimetals. The calculated band structures for a set of selected Weyl nodes in the vicinity of the Fermi level are shown in Fig. 4. According to our calculations, the tilt of the linear band crossings is small and preserves an elliptical Dirac cone with the point-like Fermi surface. As shown in previous theoretical studies [74], tilting of the Dirac cone, caused by the time-reversal symmetry breaking, can result in a one-dimensional chiral anomaly, which in turn shows linear negative MR. Therefore, our electronic structure calculations suggest that the observed linear negative MR below the quantum limit originates from the one-dimensional chiral anomaly as a distinct feature of time-reversal symmetry broken (e.g., magnetic) Weyl semimetals.

In principle, the chiral anomaly induced negative MR should be accompanied by SdH oscillations originating from the bulk three-dimensional Weyl fermions, because the chiral anomaly stems from the Landau quantization of the bulk three-dimensional Weyl fermions. However, oscillations originating from the bulk three-dimensional Weyl fermions were not clear in the SrRuO$_3$ samples studied here, as will be described in the Sec. III C. Since the Landau levels could be broadened by impurity scattering, disorder in the crystals, if any, readily hampers the observation of clear quantum oscillations [80,81]. In fact, in Dirac semimetal Na$_3$Bi and zero-gap semiconductor GdPtBi, chiral-anomaly-induced MR without quantum oscillations of the bulk three dimensional Weyl fermions has been reported [82]. Accordingly, further improvement in the crystallinity of SrRuO$_3$ is required to observe clear SdH oscillations of the bulk three-dimensional Weyl fermions, which will allow for detailed characterization of the Fermi surface. For example, using a substrate having smaller lattice mismatch with SrRuO$_3$ appears to be a promising approach because SrRuO$_3$ film on such a substrate has fewer defects owing to the smaller lattice mismatch [83].

### C. High-mobility two-dimensional carriers

To study the dimensionality of the SdH oscillations, we investigated their angular dependence in the SrRuO$_3$ film with $t = 10$ nm (Fig. 5). The SdH oscillations are observed from $\theta = 90°$ to $30°$, but not for $\theta$ ranging from $20°$ to $0°$ [Fig. 5(a)]. This behavior of weakened oscillations with decreasing $\theta$ down to the in-plane direction is a typical feature of two-dimensional carriers [6,12,84]. As shown in Fig. 5(b), the peak frequency $F$ of the FFT spectra gradually shifts from ~25 T to a high frequency with $B$ rotated from $\theta = 90°$ to 30°. Here, $k_F$ is the Fermi wave number. The frequency is well described by a $1/\cos(90°-\theta)$ dependence [Fig. 5(c)], indicating that the field component perpendicular to the surface is relevant for the cyclotron orbit. This angular dependence is a hallmark of two-dimensional carriers [6,10,12,84]. As shown in Figs. 5(d) and 5(e), we also measured the temperature dependence of the SdH oscillations and estimated the effective cyclotron mass $m^*$ according to the Lifshitz-Kosevich theory at each angle as follows [9,27,85,86]:

$$\Delta\sigma_{xx} \propto \frac{T}{\sinh(\alpha T)}, \quad (2)$$

where $\alpha = \frac{2\pi^2 k_B m^*}{\hbar e \bar{B}}$, $k_B$ is the Boltzmann coefficient, and $\bar{B}$ is defined as the average inverse field of the FFT interval. The estimated $m^*$ at $\theta = 90°$ is $0.25m_0$ ($m_0$: the free electron mass in a vacuum), and it also shows two-dimensional $1/\cos(90°-\theta)$ dependence [Fig. 5(c)] [6]. We also determined the Dingle temperature $T_D$ from the Lifshitz-Kosevish theory (see the Sec. III of the Supplemental Material [62]) and obtained the quantum mobility $\mu_q = e\hbar/(2\pi k_B m^* T_D) = 3.5 \times 10^3$ cm$^2$/Vs. These results confirm the existence of the high-mobility two-dimensional carriers with a light cyclotron mass in the SrRuO$_3$ film with $t = 10$ nm. The $F$ and $m^*$ values ($F \sim 25$ T and $m^* = 0.25m_0$) of the SrRuO$_3$ film with $t = 10$ nm are, respectively, the same as and comparable to those of the main component of the SdH oscillations in the SrRuO$_3$ film with $t = 63$ nm ($F \sim 26$ T and $m^* = 0.35m_0$) reported previously [27]. Notably, the $\mu_q$ value of the SrRuO$_3$ film with $t = 10$ nm is about one third of that in the SrRuO$_3$ film with $t = 63$ nm ($\mu_q = 9.6 \times 10^3$ cm$^2$/Vs). The reduced $\mu_q$ value in the former film suggests that disorder near the interface between SrRuO$_3$ and SrTiO$_3$ substrate is the dominant scattering mechanism, which shortens the quantum life time of the high mobility two-dimensional carriers in the SrRuO$_3$ film with $t = 10$ nm. This disorder may be the origin of the short-range scattering mechanism postulated in the theory of the saturation of the negative MR in the quantum limit [75,78,79].

As confirmed in Fig. 5(a), the SdH oscillation amplitude up to 14 T (= 0.071 T$^{-1}$) becomes small at lower angles, and it is difficult to follow the frequency. To further scrutinize the quantum oscillations, especially at lower angles ($\theta \leq 30°$), we analyzed the SdH oscillations using the mid-pulse magnet up to 52 T in the SrRuO$_3$ film with $t = 10$ nm (Fig. 6). Except for 5.3°, the SdH oscillations can be clearly seen at each angle. At $\theta = 90.8°$ and 80.8°, the SdH oscillations are distinct between the two $1/B$ ranges [Fig. 6(a)]: the high $1/B$ range from 0.035 to 0.1 T$^{-1}$ and the low $1/B$ range from 0.018 to 0.035 T$^{-1}$. The SdH oscillation at $\theta = 90.8°$ in the high $1/B$ range has the frequency $F \sim 25$ T, which corresponds to the Weyl fermions with the high $\mu_q$ and light $m^*$ observed in Fig. 6(a). The other oscillation has a high frequency of $F \sim 300$ T and the cyclotron mass $m^*$ of 2.4 $m_0$ (see the Sec. IV of the Supplemental Material [62]). These $F$ and $m^*$ values are consistent with those of the trivial Ru 4$d$ band that crosses the $E_F$, which were reported in an early de Haas–van Alphen measurement [87] and an SdH measurement using an AC analog lock-in technique at 0.1 K up to 14 T [27]. Since our focus is not on the trivial Ru 4$d$ orbit in the low 1/$B$ range but on the Weyl fermions, the data in the low 1/$B$ range was excluded from the raw data for $\theta$ = 90.8° and 80.8° for the FFT analysis. As in the case of the PPMS measurements, the SdH peak frequency gradually shifts to a high frequency with decreasing $\theta$ from 90.8° to 15.6° in Figs. 6(b) and 6(c). In Fig. 6(a), the quantum limit, which is the inverse value of the FFT frequency, is indicated by black arrows. Since the FFT frequencies at high angles ($\theta \geq$ 61.6°) are below the highest applied magnetic field of 52 T, we can observe the quantum limit as evidenced by the disappearance of the SdH oscillations of the two-dimensional carriers in the high-$B$ region above the SdH frequencies. The frequency obtained by the high field measurements is also well fitted by 1/cos(90°−$\theta$) [Fig. 6(c)], as is the case in the PPMS measurements.

We also measured the SdH oscillations for the Weyl fermions in the SrRuO$_3$ film with $t$ = 10 nm at 0.7 K using a long-pulse magnet up to 40 T (Fig. S2 [62]), which allows for a more accurate measurement of resistivity because measurement time is longer than that of the mid-pulse magnet up to 52 T. Since the SdH oscillations are clearly observed from low magnetic fields by improved measurement accuracy, the spectra of the Fourier transform have sharper peaks than those obtained by the PPMS and the mid-pulse magnet measurements, as shown in Fig. S2(b) [62]. All the results from PPMS, mid-, and long-pulse magnet measurements converge and are well fitted with an angular-dependence curve for two-dimensional carriers [Fig. 6(c)], strongly indicating the existence of the two-dimensional Weyl fermions in the SrRuO$_3$ film with $t$ = 10 nm. Altogether, the existence of high-mobility two-dimensional carriers, which had never been observed in ferromagnets, is established in the epitaxial ferromagnetic oxide SrRuO$_3$.

**D. Origin of the high-mobility two-dimensional carriers: surface Fermi arcs**

Possible origins of the high-mobility two-dimensional carriers in Weyl semimetals are the surface Fermi arcs (Fig. 8) and the quantum confinement of the three-dimensional Weyl fermions [8–15,88]. Since SrRuO$_3$ has the pseudocubic structure [83], the pseudocubic [001] (perpendicular to the film) and [100] directions (parallel to the film) are equivalent. This means that the two-dimensional carriers do not come from a highly anisotropic Fermi pocket. The SdH oscillations mediated by the surface Fermi arcs and the quantum confinement appear when the film is thinner than the mean free path and the Fermi wavelength of carriers, respectively [6,12,88]. The mean free path $\hbar^2 k_F/(2\pi k_B m^* T_D)$ and the Fermi wavelength $2\pi/k_F$ estimated from the $k_F$, $m^*$, $T_D$ values obtained by the SdH oscillations of the high-mobility two-dimensional carriers for the SrRuO$_3$ film with $t$ = 10 nm in Fig. 5(a) are ~65.4 and ~22.4 nm, respectively. Hence, the film thickness of 10 nm satisfies both conditions. In the SrRuO$_3$ film with $t$ = 63 nm, the mean free path and the Fermi wavelength of the main component of the SdH oscillations are ~179 and ~22.4 nm, respectively [27]. To determine the origin of the two-dimensional quantum transport, we also studied the angular dependence of the SdH oscillations in the SrRuO$_3$ film with $t$ = 63 nm, which is thinner than the mean free path but thicker than the Fermi wavelength. We note that, in Dirac semimetal Cd$_3$As$_2$, suppression of backscattering, which results in a transport lifetime $10^4$ times longer than the quantum lifetime, has been reported [89]. Therefore, the mean free path estimated by quantum lifetime may also be underestimated in SrRuO$_3$. Figure 7 shows the angular dependence of the SdH oscillations in the SrRuO$_3$ film with $t$ = 63 nm. As in the case of the SrRuO$_3$ thin film with $t$ = 10 nm, the peak frequency of the FFT spectra gradually shifts to a higher frequency with decreasing $\theta$ from 90° to 10° in Fig. 7(b). The frequency also shows the two-dimensional 1/cos(90°−$\theta$) dependence, with the exception of that at $\theta$ = 0° [Fig. 7(c)]. This result suggests that the high mobility two-dimensional carriers come from the surface Fermi arcs, not from the quantum confinement effect. The SdH peak frequency of 64 T at $\theta$ = 0° is outside the above two-dimensional trend, indicating that these SdH oscillations are different from those of the high mobility two-dimensional carriers. Additionally, for $\theta \geq 40°$, a small FFT peak is also discernible, e.g., at ~44 T for $\theta$ = 90° [Fig. 7(b)]. These peaks with the weak angular dependence possibly come from the orbit of bulk three-dimensional Weyl fermions. Since the SdH oscillations from the surface Fermi arcs are connected via the Landau quantization of the bulk three-dimensional Weyl fermions [88], further improvement in the crystallinity of SrRuO$_3$ will allow for detailed characterization of the SdH of the bulk Weyl fermions and their relevance to the surface Fermi arcs.

The thickness-dependent phase shift of the SdH oscillations will provide essential evidence for the non-trivial nature of the SdH oscillations mediated by the surface Fermi arcs. According to the semiclassical analysis of the closed magnetic orbit, called Weyl orbit, formed by connecting the surface Fermi arcs via the Weyl nodes in bulk [10,88], the thickness $t$ dependence of the position of the $N$th conductivity maximum $B_N$ is expressed by

$$B_N^{-1} = e k_0^{-1} \left( \frac{(N+\gamma)\pi v_F}{E_F} - \frac{t}{\hbar} \right), \quad (3)$$

where $e$, $k_0$, $\gamma$, $v_F$, $E_F$, and $\hbar$ are the elementary charge, the length of the surface Fermi arc in reciprocal space, the sum of a constant quantum offset and the Berry phase, the Fermi velocity, the Fermi energy, and the Dirac constant. Eq. (3) describes that the inverse of the SdH oscillation frequency $1/F = B_N^{-1} - B_{N-1}^{-1}$ and the thickness-dependent phase shift $\beta(t)$ of the SdH oscillation are given by $e k_0^{-1} \left( \frac{\pi v_F}{E_F} \right)$ and $\frac{E_F t}{\pi v_F \hbar}$, respectively. Thus, the SdH oscillation originating from Weyl orbits can be expressed by

$$\Delta \sigma_{xx} \propto \cos \left[ 2\pi \left( \frac{F}{B} + \gamma + \beta(t) \right) \right] \quad (4)$$

It is apparent from these equations that the phase shift $\beta(t)$ of the SdH oscillations mediated by the surface Fermi arcs has a linear relation with $t$. To examine this non-trivial phase shift, we carried out the LL fan diagram analyses for the conductivity maximum of the SdH oscillations with various thicknesses. Fig. 9 shows the background-subtracted SdH oscillations at 2 K with $B$ (3.3 T < $B$ < 14 T) applied in the out-of-plane [001] direction of the SrTiO$_3$ substrate for the SrRuO$_3$ films with $t$ = 10, 20, 40, and 60 nm. The FFT frequencies $F$ (= 27, 26, 25, and 34 T for $t$ = 10, 20, 40, and 60 nm, respectively) are determined from the peak positions of the FFT spectra. We extracted the magnetic field $B_N$ of the $N$th conductivity maximum, and estimated the phase shift $\gamma + \beta(t)$ from the LL fan diagram for the conductivity maximum with a fixed slope of $F$ (Fig. 10). As expected from eqs. (3) and (4), we can see the linear relation between the $\beta(t)$ and $t$ (Fig. 11) as a distinguished non-trivial nature of the SdH oscillations mediated by the surface Fermi arcs. The linear thickness dependence of the shift of the SdH oscillations verifies that the high-mobility two-dimensional carriers come from the surface Fermi arcs (Fig. 8), not from the trivial origins such as quantum confinement effect, trivial 2D gas on the surface of the film, anisotropic bulk Fermi surface, etc.

From the $1/F = ek_0^{-1}\left(\frac{\pi v_F}{E_F}\right)$ value and the slope of the phase shift $\frac{E_F}{\pi v_F \hbar}$ of the SdH oscillation of the Weyl orbit, we can determine the $\frac{E_F}{v_F}$ and $k_0$ values. The estimated $\frac{E_F}{v_F}$ and $k_0$ values are 2.92 ×10$^{-8}$ eVs/m and 2.8 nm$^{-1}$, respectively. Since the length of the surface Fermi arc is not much smaller than the size of the pseudocubic Brillouin zone $\frac{2\pi}{a}$ = 15.98 nm$^{-1}$ of SrRuO$_3$, our findings will stimulate a challenge for its observation by angle-resolved photoemission spectroscopy. In addition, we can determine the $\gamma$ value from the extrapolation of the $t$ dependence of the $\gamma + \beta(t)$ to $t$ = 0. The obtained $\gamma$ value of 0.016 means that a constant quantum offset and the Berry phase obtained in the Weyl orbit are almost canceling out each other.

### IV. SUMMARY

In summary, we have systematically studied quantum oscillations observed in high-quality SrRuO$_3$ films and successfully detected high-mobility two-dimensional carriers with light cyclotron mass, which can be attributed to the surface Fermi arcs. We also observed the saturation of the negative MR in the quantum limit, which is strong evidence of the chiral anomaly. Our findings provide a new foundation for studying exotic quantum transport phenomena in magnetic Weyl semimetals embedded in oxide heterostructures. The emergence of the hitherto hidden two-dimensional Weyl states in a ferromagnet is an essential step toward realizing the novel quantum Hall effect, which should be controllable through magnetization switching.

**Data Availability**
The data that support the findings of this study are available from the corresponding author upon reasonable request.

**Acknowledgements**
S.K.T. acknowledges the support from the Japan Society for the Promotion of Science (JSPS) Fellowships for Young Scientists.

**Author Contributions**
Y.K.W. conceived the idea, designed the experiments, and directed and supervised the project. Y.K.W. and Y.Kro. grew the samples. S.K.T. and Y.K.W. carried out the sample characterizations. S.K.T., Y.K.W., and H.I. fabricated the Hall bar structures. S.K.T., Y.K.W., and K.T carried out the magnetotransport measurements. S.K.T., Y.K.W, T.N., and Y.Koh. performed the high field magnetotransport measurements. S.K.T. and Y.K.W analyzed and interpreted the data. S.A.N. and H.D. carried out the electronic-structure calculations. S.K.T. and Y.K.W. co-wrote the paper with input from all authors.

**Competing interests**
The authors declare no competing interests.

**Figures and figure captions**

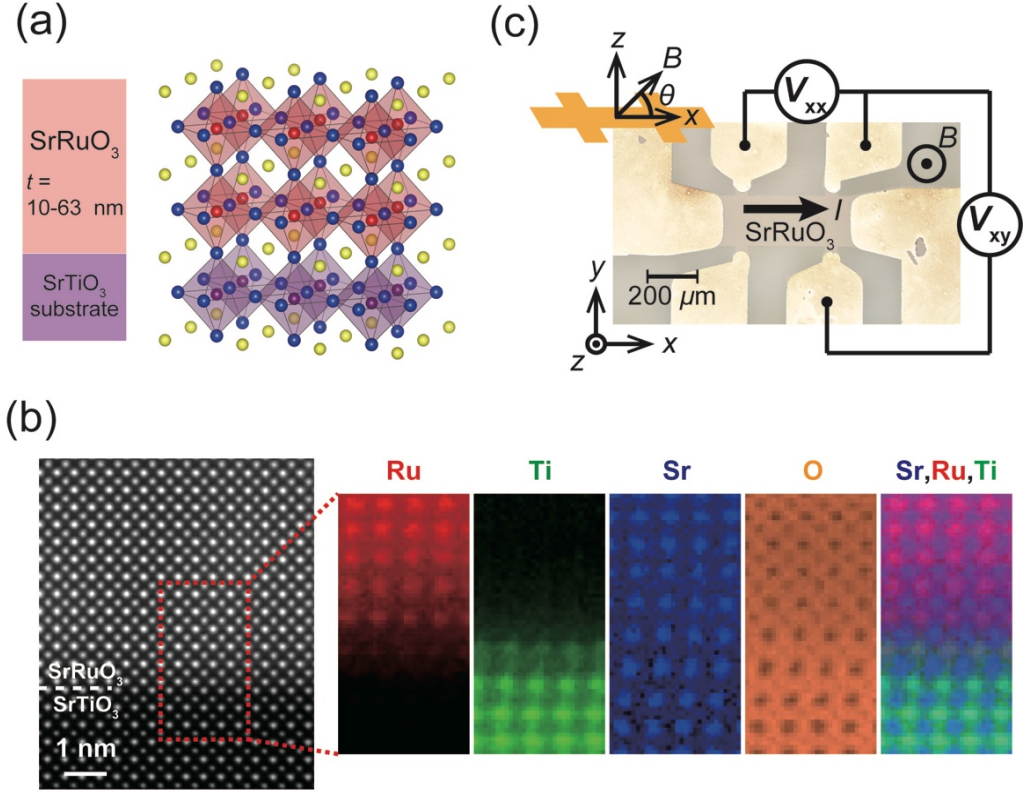

FIG. 1. (a) Schematic of the sample and crystal structures of the SrRuO₃ films on a SrTiO₃ substrate [46]. In the schematic crystal image, yellow, blue, red, and purple spheres indicate strontium, oxygen, ruthenium, and titanium, respectively. (b) Cross-sectional high-angle annular dark field scanning transmission electron microscopy (HAADF-STEM) image of the SrRuO₃ film with $t = 63$ nm along the [100] axis of the SrTiO₃ substrate. Electron energy loss spectroscopy (EELS)-STEM images (from left to right) for the Ru-$M_{4,5}$- (red), Ti-$L_{2,3}$- (green), Sr-$M_3$- (blue), O-$K$-edge (orange), and a color overlay of the EELS-STEM images for Sr, Ru and Ti. (c) Optical microscope image of a Hall bar device. The current flows along the $x$ direction, and the longitudinal and Hall voltage were measured using the $V_{xx}$ and $V_{xy}$ terminals. The upper left inset shows the arrangement of the Hall bar device and the magnetic field $B$.

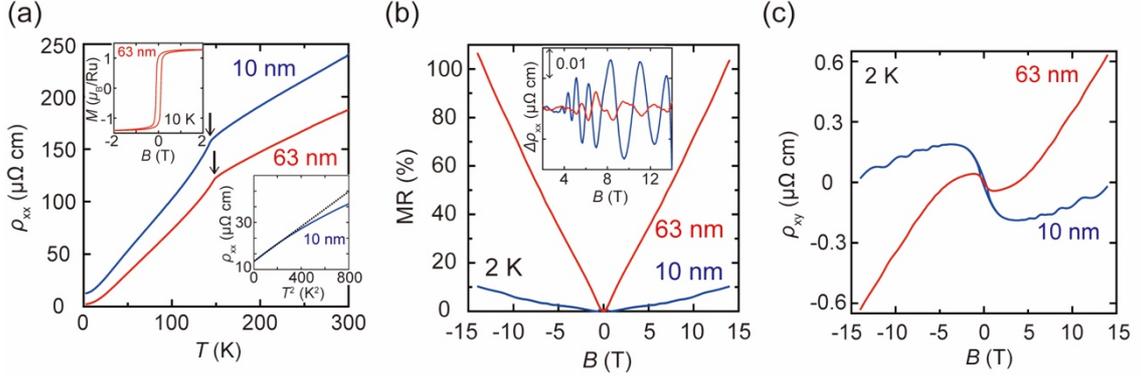

FIG. 2. (a) Temperature dependence of longitudinal resistivity $\rho_{xx}$ for the SrRuO$_3$ films with $t$ = 10 and 63 nm at zero magnetic field. The arrows indicate the ferromagnetic transition temperatures. The left inset shows the magnetization vs. magnetic field curve for the SrRuO$_3$ film with $t$ = 63 nm at 10 K with $B$ applied in the out-of-plane [001] direction of the SrTiO$_3$ substrate. The right inset shows the $\rho_{xx}$ vs. $T^2$ plot with the linear fitting (black dashed line). (b) Thickness $t$ dependence of MR $(\rho_{xx}(B)-\rho_{xx}(0\,\text{T}))/\rho_{xx}(0\,\text{T})$ at 2 K with $B$ applied in the out-of-plane [001] direction of the SrTiO$_3$ substrate ($\theta$ = 90°). The inset shows the oscillation components obtained by subtracting a polynomial function up to the eighth order. (c) Thickness $t$ dependence of transverse resistivity $\rho_{xy}$ at 2 K with $B$ applied in the out-of-plane [001] direction of the SrTiO$_3$ substrate ($\theta$ = 90°).

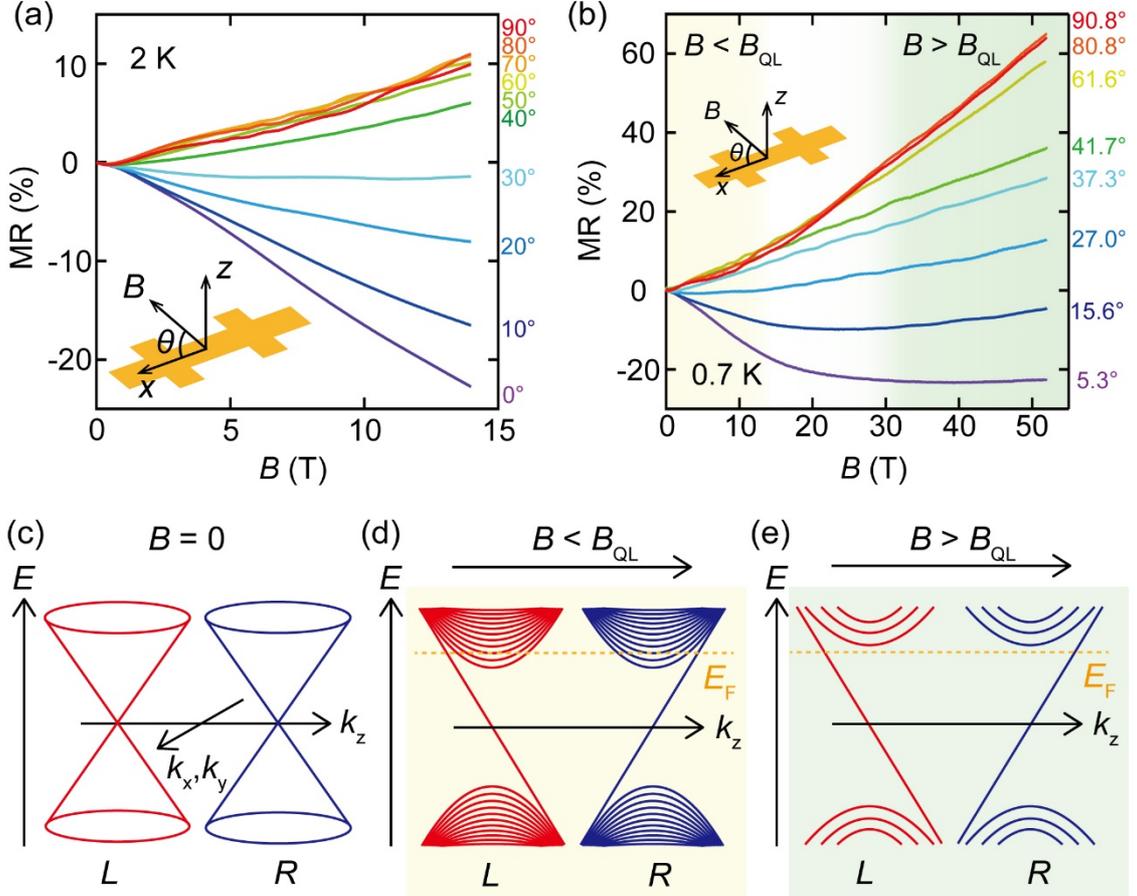

FIG. 3. (a) Angular dependence of MR $(\rho_{xx}(B)-\rho_{xx}(0\text{ T}))/\rho_{xx}(0\text{ T})$ for the SrRuO$_3$ film with $t = 10$ nm at 2 K, measured by the PPMS up to 14 T. (b) Angular dependence of MR $(\rho_{xx}(B)-\rho_{xx}(0\text{ T}))/\rho_{xx}(0\text{ T})$ for the SrRuO$_3$ film with $t = 10$ nm at 0.7 K, measured by the mid-pulse magnet up to 52 T. (c) Schematic diagram of a pair of Weyl nodes with opposite chiralities (L and R) with $B = 0$. (d), (e) Schematic diagram of Landau levels of a pair of Weyl nodes with $B < B_{QL}$ and $B > B_{QL}$, respectively, where $B_{QL}$ is the magnetic field at which the quantum limit is reached. In (b), beige and green areas indicate $B < B_{QL}$ and $B > B_{QL}$ range, respectively.

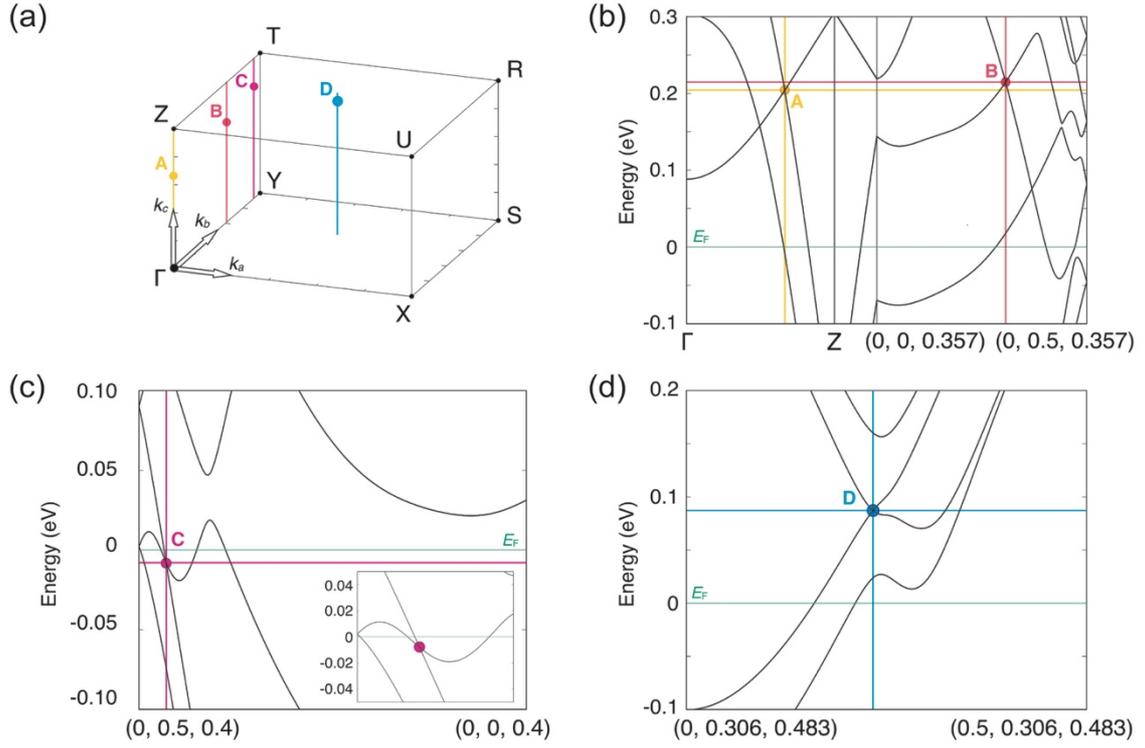

FIG. 4. Selected Weyl nodes (A-D) in the vicinity of the Fermi level in orthorhombic SrRuO$_3$. (a) Position of the Weyl nodes in the irreducible part of the Brillouin zone. Vertical yellow, red, pink, and blue lines indicate the in-plane (Γ–X–S–Y plane) positions of the Weyl nodes for easy viewing. The Weyl nodes are given in fractional coordinates as A = (0, 0, 0.332), B = (0, 0.307, 0.357), C = (0, 0.464, 0.4), and D = (0.233, 0.306, 0.483). (b)-(c) Band structures for the ferromagnetic ground state without spin-orbit coupling as obtained from GGA + $U$ calculations with $U$ = 2.6 eV and $J$ = 0.6 eV. In (b)-(d), the intersection of the straight lines represents the Weyl nodes. $E_F$ stands for the Fermi level.

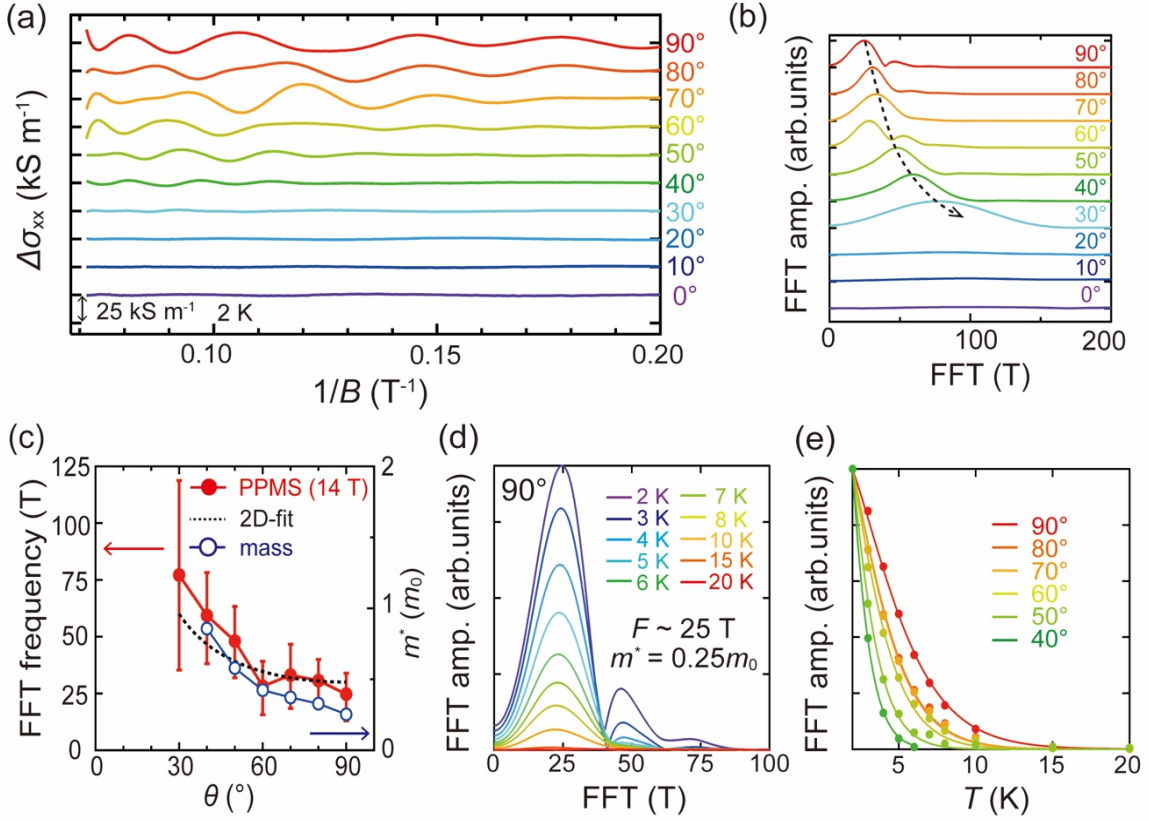

FIG. 5. (a) Angular dependence of the background-subtracted SdH oscillations at 2 K with $B$ (5 T < $B$ < 14 T) for the SrRuO$_3$ film with $t$ = 10 nm. (b) Angular dependence of the FFT spectra of SdH oscillations at 2 K for the film with $t$ = 10 nm. The dashed line with an arrow is a guide for the peak shift. (c) Angular dependence of the FFT frequency (red filled circles) and the cyclotron mass (blue open circles). The dashed line is the fitting of a two-dimensional angular dependence (~1/cos(90°−$\theta$)). (d) Temperature dependence of the Fourier transform spectra of SdH oscillations at $\theta$ = 90° for the SrRuO$_3$ film with $t$ = 10 nm. (e) Temperature dependence of the FFT amplitude from $\theta$ = 90° to 40° for mass estimations according to the Lifshitz-Kosevich theory.

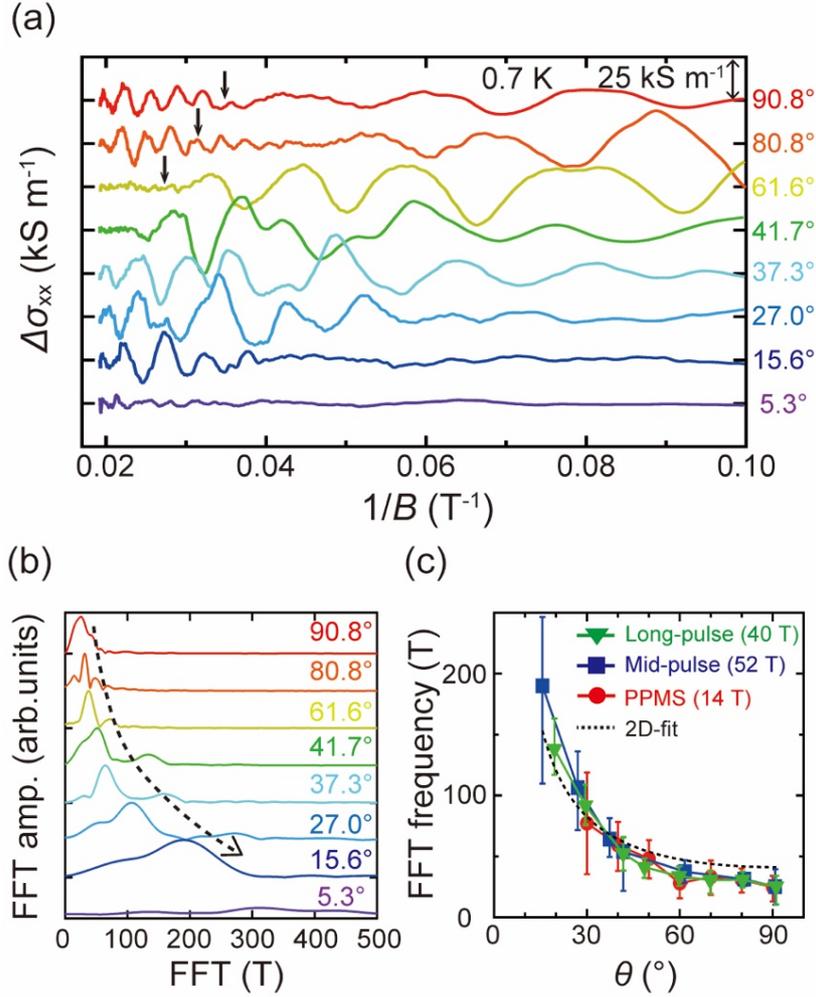

FIG. 6. (a) Angular dependence of the background-subtracted SdH oscillations at 0.7 K with $B$ (10 T < $B$ < 52 T) for the SrRuO$_3$ film with $t$ = 10 nm, measured by the mid-pulse magnet. These spectra were smoothed (see the Sec. V of the Supplemental Material [62]). The arrows indicate the quantum limit, which is the inverse value of the FFT frequency in (c). (b) Angular dependence of the FFT spectra of SdH oscillations at 0.7 K for the film with $t$ = 10 nm. The dashed line with an arrow is a guide for the peak shift. (c) Angular dependence of the FFT frequency. Here, the data obtained by the long-pulse magnet are also shown. The dashed line is the fitting of a two-dimensional angular dependence (~$1/\cos(90-\theta)$).

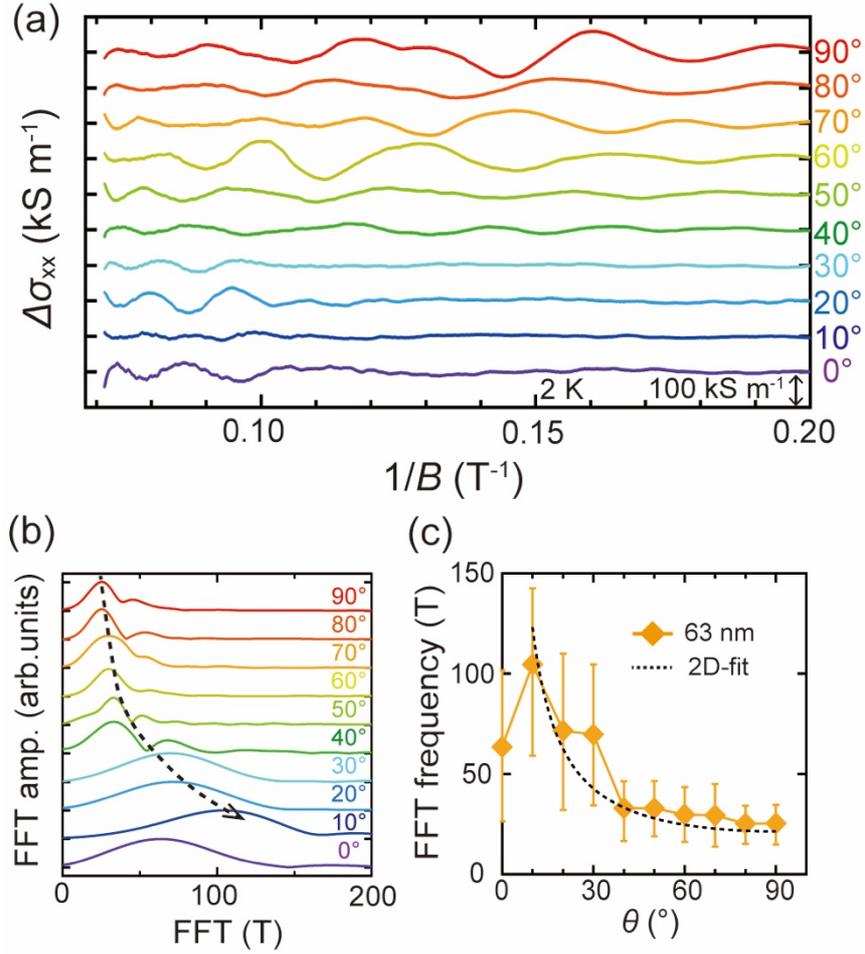

FIG. 7. (a) Angular dependence of the background-subtracted SdH oscillations at 2 K with $B$ (5 T < $B$ < 14 T) for the SrRuO$_3$ film with $t$ = 63 nm. (b) Angular dependence of the FFT spectra of SdH oscillations at 2 K for the film with $t$ = 63 nm. The dashed line with an arrow is a guide for the peak shift. (c) Angular dependence of the FFT frequency. The dashed line is the fitting of a two-dimensional angular dependence ($\sim 1/\cos(90°-\theta)$).

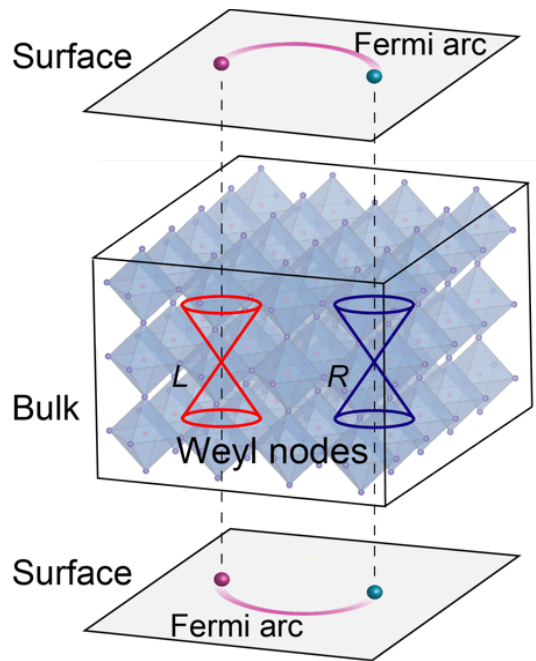

FIG. 8. Schematic diagram of a pair of Weyl nodes with opposite chiralities (*L* and *R*) and surface Fermi arcs in SrRuO$_3$.

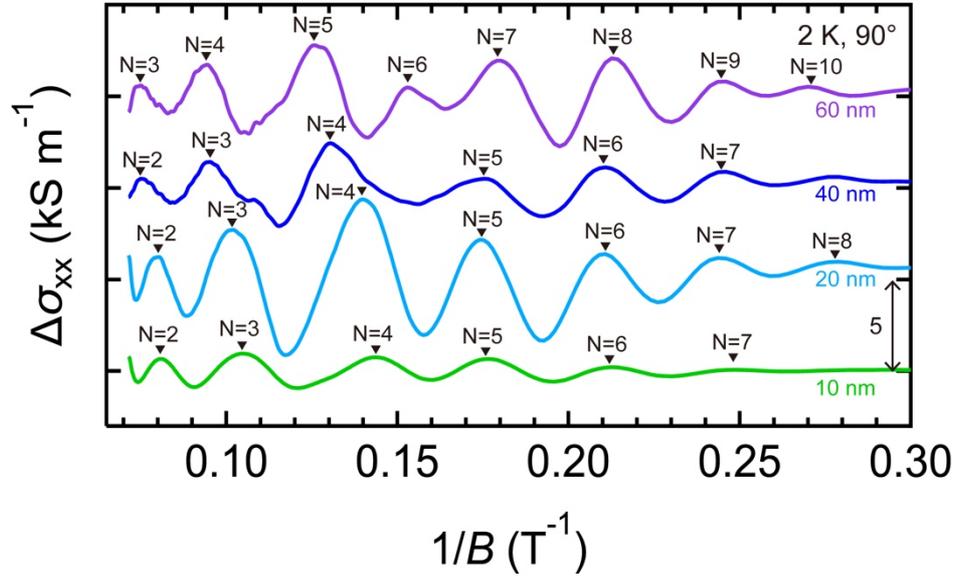

FIG. 9. Background-subtracted SdH oscillations at 2 K with $B$ (3.3 T $< B <$ 14 T) applied in the out-of-plane [001] direction of the $SrTiO_3$ substrates for the $SrRuO_3$ films with $t =$ 10, 20, 40, and 60 nm. Black triangles show peak positions.

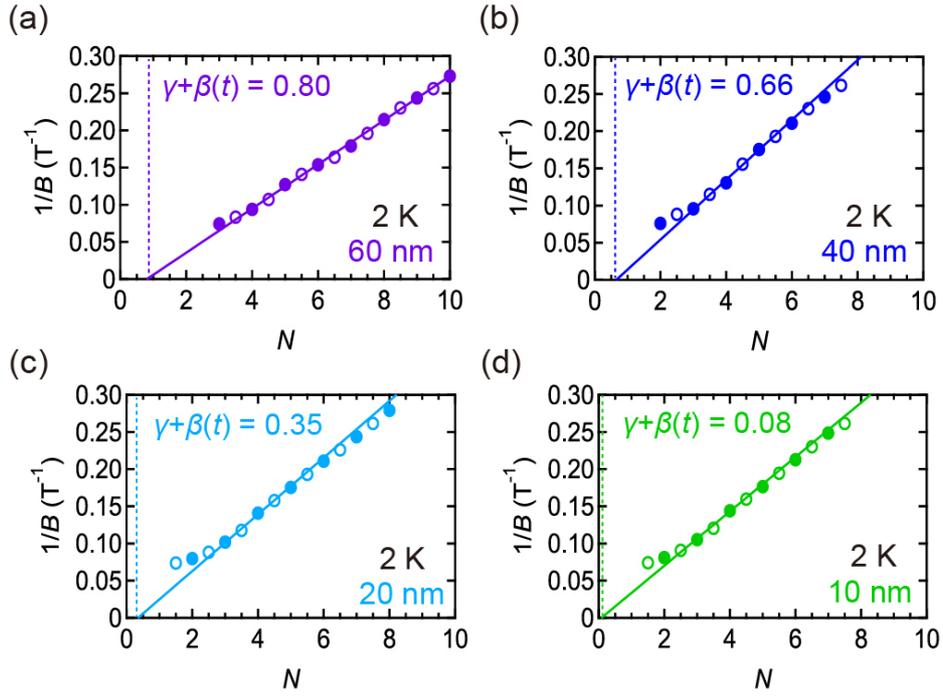

FIG. 10. LL fan diagram of the conductivity maximum for the SrRuO$_3$ films with $t$ = 10, 20, 40, and 60 nm. The fixed slopes of the linear fittings are obtained by the frequency F. $\beta$ is the phase shift in eq. (4).

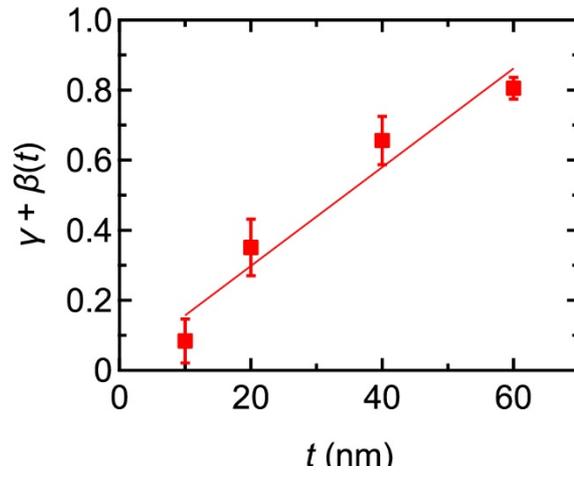

FIG. 11. Thickness $t$ dependence of the $\gamma + \beta(t)$. The red line represents the linear fitting. The error bars are defined as the standard deviation of the linear fittings in Fig. 10.

# Supplemental Material for
# High-mobility two-dimensional carriers from surface Fermi arcs in magnetic Weyl semimetal films

Shingo Kaneta-Takada,[1,2,*] Yuki K. Wakabayashi,[1,*,†] Yoshiharu Krockenberger,[1] Toshihiro Nomura,[3] Yoshimitsu Kohama,[3] Sergey A. Nikolaev,[4,5] Hena Das,[4,5] Hiroshi Irie,[1] Kosuke Takiguchi,[1,2] Shinobu Ohya,[2,6] Masaaki Tanaka,[2,7] Yoshitaka Taniyasu,[1] and Hideki Yamamoto[1]

[1]*NTT Basic Research Laboratories, NTT Corporation, Atsugi, Kanagawa 243-0198, Japan*
[2]*Department of Electrical Engineering and Information Systems, The University of Tokyo, Bunkyo, Tokyo 113-8656, Japan*
[3]*Institute for Solid State Physics, The University of Tokyo, Kashiwa, Chiba 277-8581, Japan*
[4]*Laboratory for Materials and Structures, Tokyo Institute of Technology, 4259 Nagatsuta, Midori-ku, Yokohama, Kanagawa 226-8503, Japan*
[5]*Tokyo Tech World Research Hub Initiative (WRHI), Institute of Innovative Research, Tokyo Institute of Technology, 4259 Nagatsuta, Midori-ku, Yokohama, Kanagawa 226-8503, Japan*
[6]*Institute of Engineering Innovation, The University of Tokyo, Bunkyo, Tokyo 113-8656, Japan*
[7]*Center for Spintronics Research Network (CSRN), The University of Tokyo, Bunkyo, Tokyo 113-8656, Japan*

[*]These authors contributed equally to this work.
[†]Corresponding author: yuuki.wakabayashi.we@hco.ntt.co.jp

# I. Raw conductivity $\sigma_{xx}(B)$ data for SrRuO$_3$ films with $t$ = 10 and 63 nm measured by the PPMS, mid-pulse magnet, and long-pulse magnet

Figure S1 shows the raw conductivity $\sigma_{xx}(B)$ data of the conductivity tensor for the SrRuO$_3$ films with $t$ = 10 and 63 nm. The conductivity also shows the linear $B$ dependence below the quantum limit.

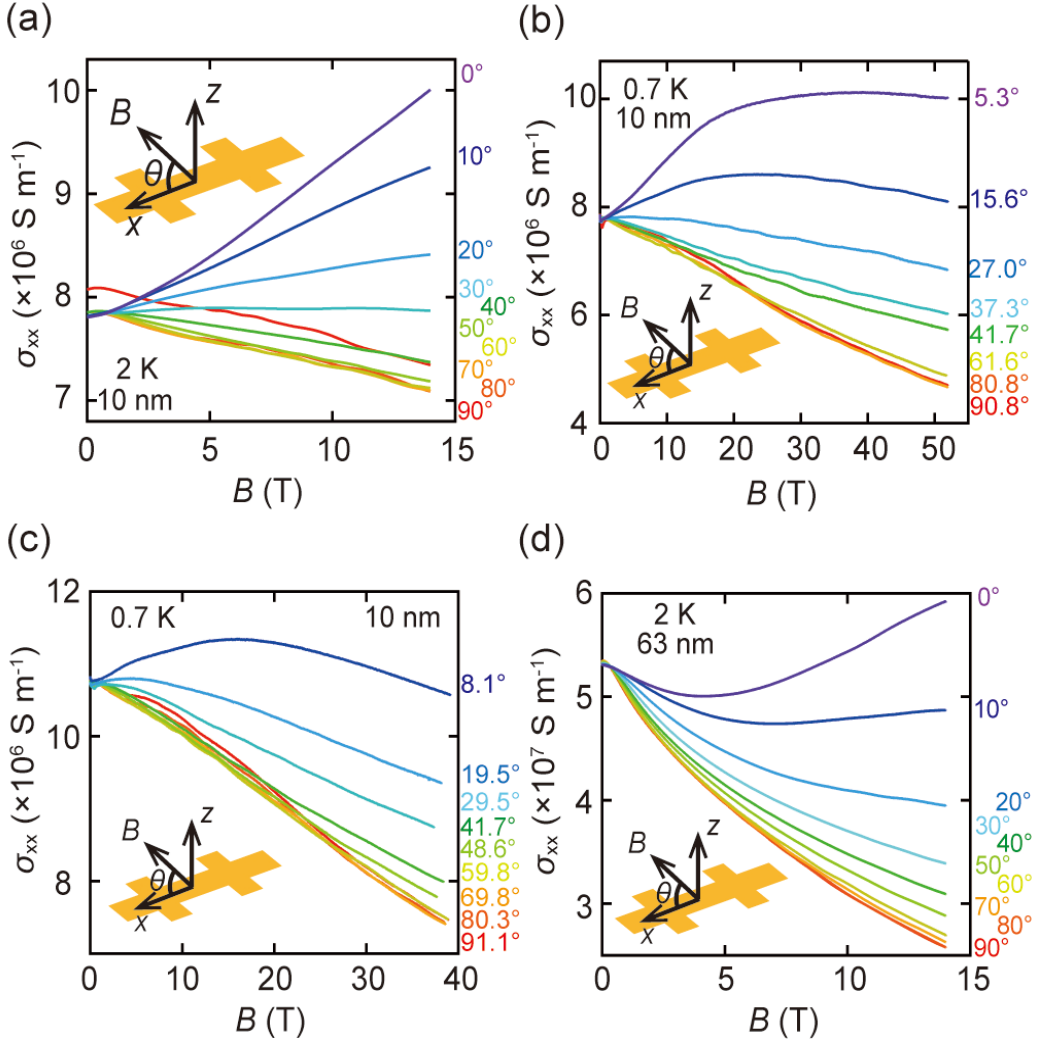

FIG. S1. (a)–(c) Angular dependence of the raw $\sigma_{xx}(B)$ for the SrRuO$_3$ film with $t$ = 10 nm measured by the PPMS up to 14 T, the mid-pulse magnet up to 52 T, and the long-pulse magnet up to 40 T, respectively. (d) Angular dependence of the raw $\sigma_{xx}(B)$ for the SrRuO$_3$ film with $t$ = 63 nm measured by the PPMS up to 14 T.

## II. High-field angle-dependent magnetotransport properties for the SrRuO$_3$ film with $t = 10$ nm using the long-pulse magnet up to 40 T

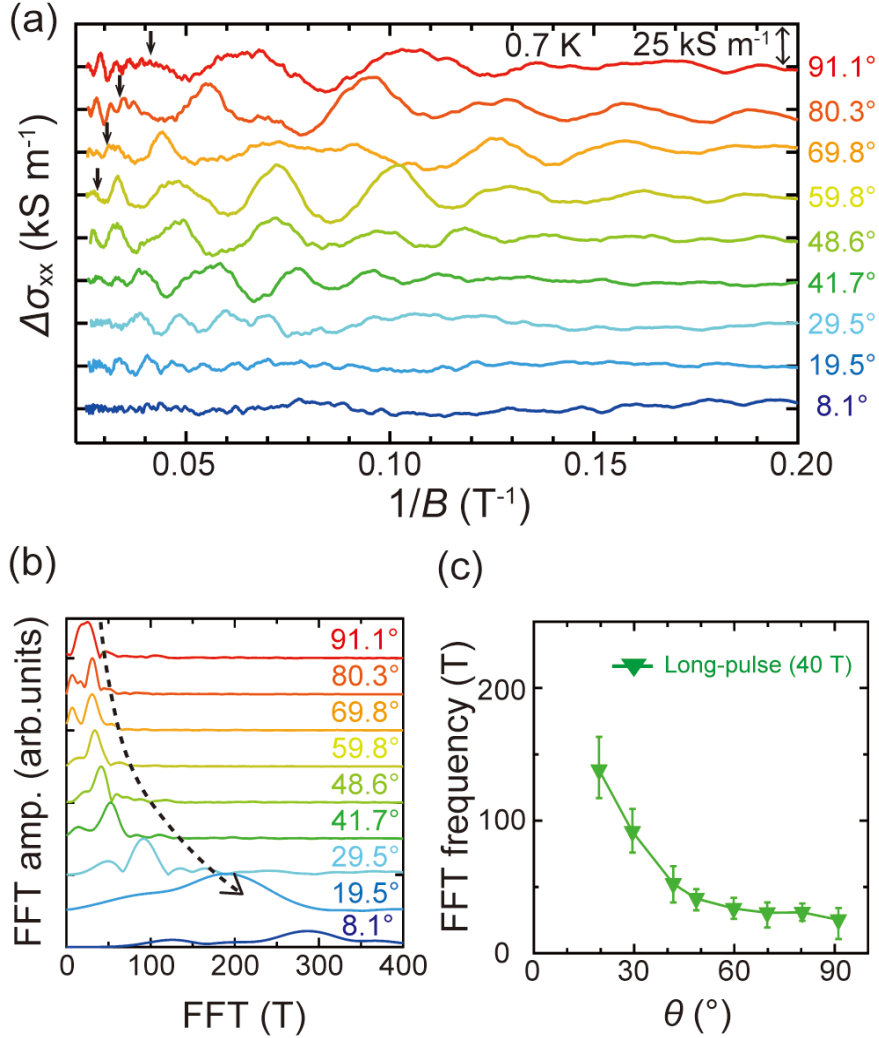

FIG. S2. (a) Angular dependence of the background-subtracted SdH oscillations at 0.7 K with $B$ (5 T < $B$ < 40 T) for the SrRuO$_3$ film with $t = 10$ nm, measured by the long-pulse magnet up to 40 T. These spectra were smoothed (see the section V of the Supplemental Material). The arrows indicate the quantum limit, which is the inverse value of the FFT frequency in (c). (b) Angular dependence of the FFT spectra of SdH oscillations at 0.7 K for the film with $t = 10$ nm. The dashed line with an arrow is a guide for the peak shift. (c) Angular dependence of the FFT frequency.

## III. Determination of Dingle temperature and Berry phase of SrRuO$_3$ film with $t$ = 10 nm from Lifshitz-Kosevich theory

We determined the Dingle temperature $T_D$ of the SrRuO$_3$ film with $t$ = 10 nm from the Lifshitz-Kosevish (LK) theory, which is described as [S1]

$$\Delta\sigma_{xx} = A \frac{X}{\sinh X} \exp\left(-\frac{2\pi^2 k_B T_D}{\hbar\omega_c}\right) \cos\left[2\pi\left(\frac{F}{B} + \gamma + \beta(t)\right)\right],$$

where $\Delta\sigma_{xx}$ is the oscillation component of the longitudinal conductivity, $A$ is the normalization factor, $X = 2\pi^2 k_B T/\hbar\omega_c$, $k_B$ is the Boltzmann constant, $\hbar$ is the reduced Planck constant, $\omega_c$ is the cyclotron frequency defined as $eB/m^*$, $m^*$ is the cyclotron mass of $0.25 m_0$ ($m_0$: electron rest mass), $T_D$ is the Dingle temperature, $F$ is the frequency of the SdH oscillation, $\gamma$ is the the sum of a constant quantum offset and the Berry phase, and $\beta(t)$ is the phase shift in eq. (4). We used the above formula to fit to the SdH oscillation data (Fig. S3). From the fitting, we obtained $T_D$ = 2.42 K. The obtained quantum mobility $\mu_q = e\hbar/(2\pi k_B m^* T_D)$ is $3.5\times10^3$ cm$^2$/Vs.

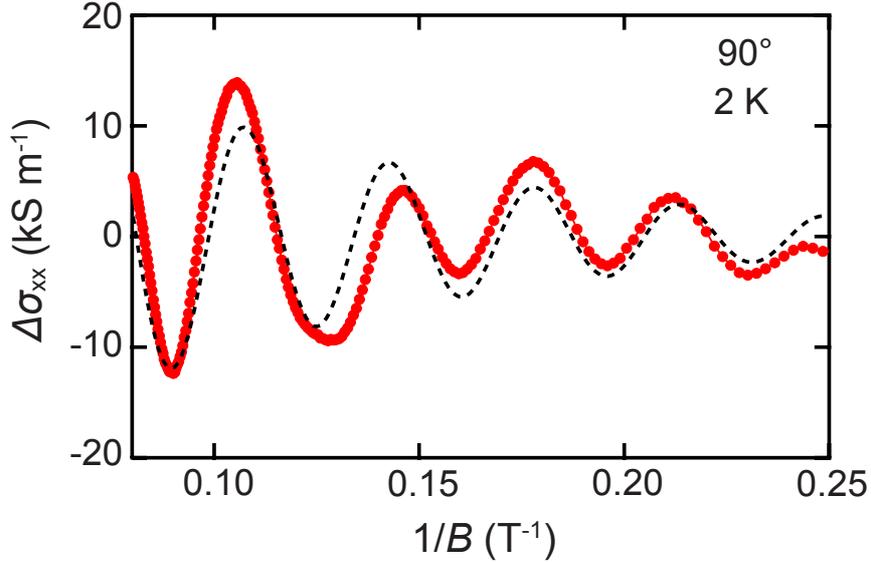

FIG. S3. Background-subtracted SdH oscillation at 2 K with $B$ (4 T < $B$ < 12.5 T) applied in the out-of-plane [001] direction of the SrTiO$_3$ substrate for the SrRuO$_3$ film with $t$ = 10 nm. The black dashed curve is the LK formula fitting with the Dingle temperature $T_D$ of 2.47 K.

**IV. Temperature dependence of SdH oscillations of trivial Ru 4*d* band**

In the main text, we focus on the SdH oscillations having a low frequency (~25 T) with $B$ applied along the out-of-plane direction to the SrTiO$_3$ substrate. Here, we analyze the SdH oscillations having high frequency (~300 T). Figs. S4(a) and S4(b) show the temperature dependence the SdH oscillations having a high frequency. The cyclotron mass $m^*$ of the high-frequency oscillation obtained from the Lifshitz-Kosevich theory is $2.4m_0$ [Fig. S4(c)]. This is consistent with that of the trivial Ru 4*d* band that crosses the $E_F$, which was reported in an earlier de Haas–van Alphen measurement [S2] and in an SdH measurement using an AC analog rock-in technique at 0.1 K up to 14 T [S3].

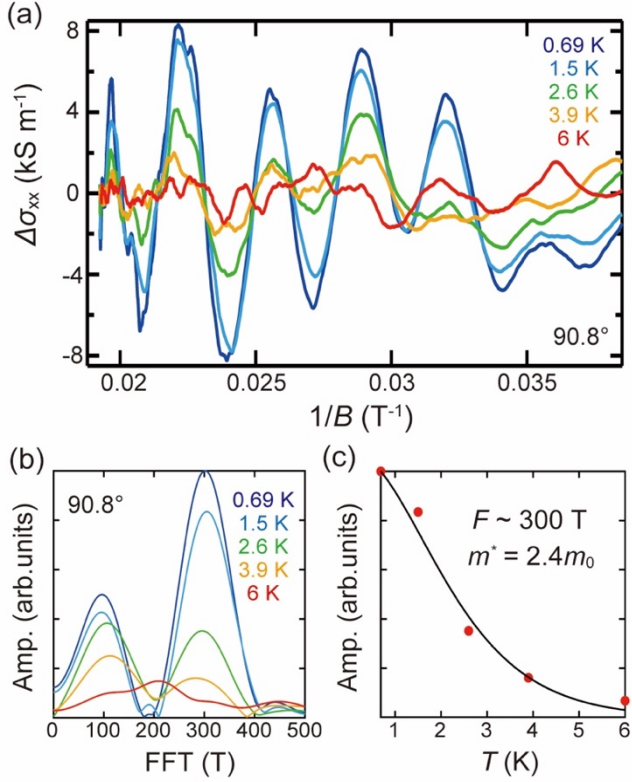

FIG. S4. (a) Temperature dependence of the background-subtracted high-frequency SdH oscillations with $F \sim 300$ T at 90.8° with $B$ (25 T < $B$ < 52 T) for the SrRuO$_3$ film with $t$ = 10 nm. (b) Temperature dependence of the FFT spectra of the SdH oscillations at $\theta$ = 90.8° for the SrRuO$_3$ film with $t$ = 10 nm. (c) Mass estimation from the Lifshitz-Kosevich theory.

## V. Angle dependence of raw data and data after smoothing of background-subtracted SdH oscillations

In the main text, we used the background-subtracted SdH oscillations data after smoothing. The raw data are shown in Figs. S5 and S6.

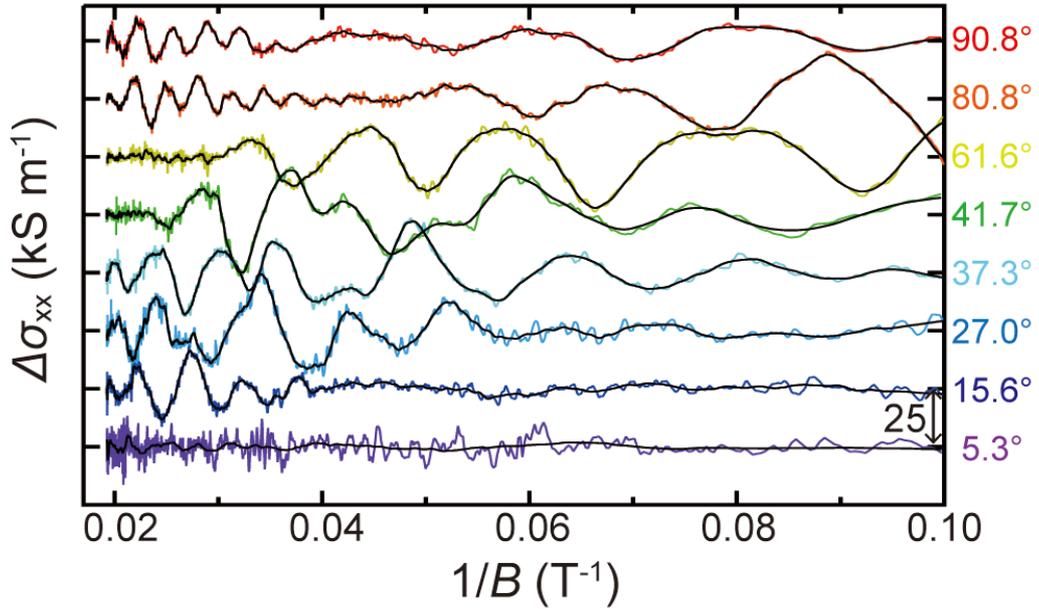

FIG. S5. Angular dependence of the background-subtracted SdH oscillations at 0.7 K with $B$ (10 T < $B$ < 52 T) for the SrRuO$_3$ film with $t$ = 10 nm measured by the mid-pulse magnet up to 52 T. The colored spectra are the raw data, and black lines are the data after smoothing.

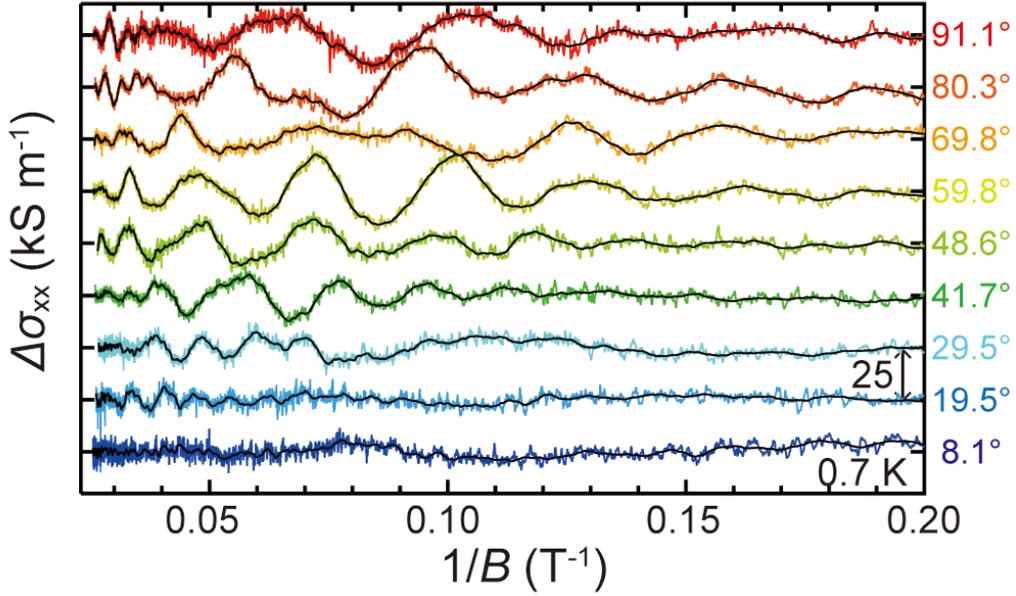

FIG. S6. Angular dependence of the background-subtracted SdH oscillations at 0.7 K with $B$ (5 T < $B$ < 40 T) for the SrRuO$_3$ film with $t$ = 10 nm measured by the long-pulse magnet up to 40 T. The colored spectra are the raw data, and black lines are the data after smoothing.